\documentclass[review]{elsarticle}
\usepackage{array}
\usepackage{longtable}
\usepackage{amsmath}
\usepackage{graphicx}
\usepackage{amssymb}
\usepackage{array}
\usepackage{float}

\journal{Chaos, Solitons and Fractals}

\newcommand{\noun}[1]{\textsc{#1}}
\providecommand{\tabularnewline}{\\}

\begin{document}

\begin{frontmatter}

\title{Emergence of coherent motion in aggregates of motile coupled maps}

\author[PIK-Potsdam,ULB]{A. Garc\'{i}a Cant\'{u} Ros\corref{cor1}}
\ead{anselmo@pik-potsdam.de}

\author[ULB]{Ch.G. Antonopoulos}
\ead{cantonop@ulb.ac.be}

\author[ULB]{V. Basios}
\ead{vbasios@ulb.ac.be}

\address[PIK-Potsdam]{Potsdam Institute for Climate Impact Research, 14412, Potsdam, Germany}

\address[ULB]{Interdisciplinary Center for Nonlinear Phenomena and Complex Systems (CENOLI), Service de Physique des Syst\`{e}mes Complexes et M\'{e}canique Statistique, Universit\'{e} Libre de Bruxelles, 1050, Brussels, Belgium}

\cortext[cor1]{Corresponding author}

\begin{abstract}
In this paper we study the emergence of coherence in collective motion described by a system of interacting motiles endowed with an inner, adaptative, steering mechanism. By means of a nonlinear parametric coupling, the system elements are able to swing along the route to chaos. Thereby, each motile can display different types of behavior, i.e. from ordered to fully erratic motion, accordingly with its surrounding conditions. The appearance of patterns of collective motion is shown to be related to the emergence of interparticle synchronization and the degree of coherence of motion is quantified by means of a graph representation. The effects related to the density of particles and to interparticle distances are explored. It is shown that the higher degrees of coherence and group cohesion are attained when the system elements display a combination of ordered and chaotic behaviors, which emerges from a collective self-organization process.
\end{abstract}


\end{frontmatter}

\section{Introduction}

Complex motion modes of collectives as a result of their constituent
interacting entities occurs almost ubiquitously in nature and over
the last decades it has provided a common ground for cross-disciplinary
investigations among Physics, Biology and Mathematics. The range of
applications of such studies is indeed extensive \cite{DeneubourgBook,Foundations2007,Nicolis99,IG08,StatPhysBirds2007,science99,Kaneko94}.
As a matter of fact, coherent patterns of collective motion found
in distinct families of biological species such as fish schools, flocks
of birds, swarms of insects and even colonies of bacteria \cite{science99,StatPhysBirds2007,Chowdhury2005,DeneubourgBook,IG08,Matteo08,Parrish02,WoodAckland2007}
have also been detected in granular matter systems, self-propelled
particles with inelastic collisions and active Brownian particles
in autonomous-motor groups \cite{AKaJMaIGK07,Dossetti09,EMRaMMMaDRC95,GGaHC04,Miguel03,Parravano08,SFaEWaTB01,TSaKK03,UEaWE02,VIRaDBaVLKaAVZ07,WLaHZaMZCaTZ08}.
Early in the study of such collectives, modeling and simulation have
been recognized as playing a crucial role in gaining insight of the
mechanisms underlying such an emergence of global features from a
set of simple rules \cite{DenebourgGoss89,Kaneko87,Miramomdes95}.

As the interest of the scientific community in addressing such kind
of systems increases, minimal microscopic models have been recently
introduced. Most of them consider systems of many interacting elements
whose couplings, being in general nonlinear, can be either local or
global. From a purely deterministic perspective, such systems can
be represented as high dimensional dynamical systems, which can be
either discrete or continuous in their time evolution. This approach
has been mainly spearheaded by Smale and collaborators \cite{FCaSS07,FCaSS07b,SFaEWaTB01,VIRaDBaVLKaAVZ07}.
On the other hand, providing understanding of the connection between
macroscopic collective features and microscopic scale interactions
in multi-particle systems is well within the principal aims of statistical
physics. Therefore, naturally, models of self-propelled interacting
particles have provided a fertile ground for study using concepts
and tools of statistical physics. This kind of approach has been mainly
adopted in models where randomness is introduced in the dynamics of
the particles by means of a Langevin-type description \cite{WLaHZaMZCaTZ08,VIRaDBaVLKaAVZ07,UEaWE02,TSaKK03,StatPhysBirds2007,Parravano08,OJOaMRE99,MNaIDaTV07,Dossetti09,Aldana07}.
Finally, another line of investigation aims at providing descriptions
and modeling in purely probabilistic terms where interactions obey
probabilistic `rules of engagement'. Notably, effective Fokker-Plank
equations have been recently proposed for coarse-grained observables
of `agent systems' (see for example \cite{AKaJMaIGK07,DeneubourgBook,RC09}
and references therein for a more detailed presentation).

One of the earliest theoretical stochastic models of self-propelled
interacting particles was introduced by Vicsek and collaborators as
early as in 1995 \cite{MNaIDaTV07,StatPhysBirds2007}, which still
possesses seminal value because of its minimal character. In Vicsek's
model, point particles move at discrete time steps with fixed speed.
At every time step, the different particles velocities are determined
by the average of neighboring particles. In other words, Vicsek's
model is an XY model in which the `spins are actively moving' (see
\cite{HCaFGaGGaFPaFR08}). Furthermore, similarly as in ferromagnetic
spin systems, Vicsek's model exhibits a phase transition as a function
of both the particle density and the intensity of noise. For a detailed
investigation on the nature of such phase transitions we refer the
reader to \cite{MNaIDaTV07,GGaHC04,Aldana07}. Further variations
of Vicsek's model have been recently proposed to account for changing
symmetries, adding cohesion or taking into account a surrounding fluid
interacting with the particles \cite{HCaFGaGGaFPaFR08}.

Although the application of concepts stemming from statistical physics
research has led to identify some universal properties existing in
these classes of systems, such as spontaneous symmetry breaking, phase
transitions and synchronizing modes \cite{EMRaMMMaDRC95,WLaHZaMZCaTZ08,UEaWE02,OJOaMRE99,GGaHC04},
the role of the individuals' internal dynamics still remains veiled. 

Whilst the explicit consideration of inner control processes could
increase the complexity of models of interacting motiles, it is a
necessary conceptual step in developing further insights into the
mechanisms underlying the emergence of coherence of motion in biological
systems. Historically, models of group motion where particles adapt
to their environment by means of an inner steering mechanism have
been developed in the context of traffic modelling. For an overiew
of such models we refer to \cite{Bonzani2000}. In the context of
biological systems, inner states have been considered in order to model the emergence of coherent behavior in groups of 
fireflies \cite{Avila,Sumantra}. More closely related to the problem here addressed is the study of the response
and adaptation of populations of motiles to the information carried
by external `fields'. Recently, by assuming inner state dynamics,
attempts in this direction have been reported in the study of bacterial
chemotaxis \cite{Chemotaxis_Bray,Rousset_2009} and biologically inspired
collective robotics \cite{Winfield,RC09}. 

In the present work we introduce a purely deterministic model where,
in analogy to the Vicsek's class of models, particles can display
phenomenological random-like motion and exhibit a sharp change of
coherence of motion as a function of the particle density. Furthermore,
in contradistinction to Vicsek's class of models, no boundary
conditions are considered, since a feature of the group's cohesion
is that it is built by the collective dynamics `per se'. Even in the 
absence of explicit interparticle attraction, it is the coordinated effectively synchronized collective
motion that keeps the group together. In our model
every motile is endowed with an inner `steering' variable that evolves
according to an heuristic discrete-time equation. For the sake of
simplicity, such an evolution law has the structure of the logistic
map. The latter provides a suitable, well understood, combination
of chaotic and ordered behavior to account for the coherence and novelty
aspects observed in real collectives of motiles. Communication between
a particle and its environment occurs via a control parameter that
tunes its value according to the external states of the surrounding
particles. At the microscopic level the features of motiles are summarized
by the following conditions, along the lines of \cite{TSaKK03}:

\noindent ($\boldsymbol{\alpha}$) Each element has a time dependent
internal state and spatial position.

\noindent ($\boldsymbol{\beta}$) Each element is `active' in the
sense that its internal state can exhibit chaotic behavior, both in
presence and absence of interactions with other particles.

\noindent ($\boldsymbol{\gamma}$) The dynamics of the internal state
of a given element is determined by local, short range interactions,
effectuated within a neighborhood of a characteristic radius.

\noindent ($\boldsymbol{\delta}$) The interparticle interaction depends
on the internal states of the participating particles.

\noindent These general rules have been found to give rise to nontrivial
emergent behavior which can not be readily deduced from the microscopic
parameters of the system.

At the collective or `macroscopic' level, the characteristic emergent
phenomena observed in such kind of locally coupled systems are mainly
described by the notion of `clustering', either in real or in state
space. As it has been reported in \cite{TSaKK03}, distinct classes
of clustering behavior accompany, in a generic way, such a coupling:

\noindent (i) Elements forming a cluster merge in and out of the
cluster.

\noindent (ii) Elements can remain separated from neighboring clusters
but they form a bridge between distinct clusters facilitating information
flow, exchange of elements between clusters and adding cohesion.

\noindent (iii) Presence of independent clusters separated by distances
larger than the interaction, with elements rarely merging in and out
of the clusters amidst them.

\noindent (iv) Cluster - cluster interactions such as aggregation,
segregation and competitive growth between various sized clusters.

In this work, we shall focus on the description and quantification
of emergent collective motion based on a clustering index that accounts
simultaneously for both, the degree of spatial clustering and the
degree of interparticle alignment of velocities. 

The paper is organized as follows: In Section \ref{sec:Formulation of the Model},
we introduce a general formulation of the model. Next, in Section
\ref{sec:LocalandMicro} we address the possible types of stationary
behavior in the motion of an individual particle, as well as the
basic phase-locking synchronization process that results from local
interparticle pair-interactions. Section \ref{sec:The-Emergent-Global}
presents the case of the many particle system. Its typical evolution
patterns, regimes of motion, degree of synchronization and the dependency
on both, density of particles and interparticle distance, are investigated.
Finally, Section \ref{sec:Conclusions} concludes the present work
with a brief summary, discussion of the results and possible further
extentions.

\section{Formulation of the model\label{sec:Formulation of the Model}}

We consider $N$ particles, labeled through an index $i=1,2,\ldots,N$,
whose positions at time $t$ are denoted by the vectors $\{\vec{r}_{t}^{i}\}$.
They evolve on a plane (two dimensional motion) where their positions
change simultaneously at discrete time steps $\Delta t$, according
to \begin{equation}
\vec{r}_{t+\Delta t}^{i}=\vec{r}_{t}^{i}+\vec{v}_{t}^{i}\cdot\Delta t.\label{eq:pos1}\end{equation}
 Similarly as in Vicsek's original model \cite{MNaIDaTV07,StatPhysBirds2007},
we assume here that at every time step the speeds of all particles
are equal to a common constant value \begin{equation}
s=\Vert\vec{v}_{t}^{i}\Vert.\label{eq:speed}\end{equation}
 Changes in the particles' velocities occur via an inherent steering
mechanism which can be expressed in terms of a two dimensional rotation
matrix $\mathbf{T}_{t}^{i}$ as \begin{equation}
\vec{v}_{t+\Delta t}^{i}=\mathbf{T}_{t}^{i}\cdot\vec{v}_{t}^{i}.\label{eq:rot}\end{equation}
 Assuming that the motion of each particle is governed by an inner
steering process, we endow each particle $i$ with a variable $\theta_{t}^{i}$
determining, at every time step, the phase of the rotation matrix
$\mathbf{T}_{t}^{i}\mathbf{=T}(\theta_{t}^{i})$. Phases $\theta_{t}^{i}$
are assumed to take values in an interval $\left[-\Delta_{0},\Delta_{0}\right]$,
where the maximum rotation angle $\Delta_{0}$ is taken to be a small
fraction of $\pi$. Furthermore, let us consider the evolution of
the rotation phase, for each of the particles, as determined by an
equation of the general form \begin{equation}
\theta_{t+\Delta t}^{i}=\mathbf{\Phi}_{t}^{i}\left(\theta_{t}^{i};\left\{ \vec{r}_{t}^{j},\vec{v}_{t}^{j}\right\} _{j\neq i}\right).\label{eq:bigFI}\end{equation}
 Here, the function $\mathbf{\Phi}_{t}^{i}$ is introduced to model
the response of the particle $i$ to the influence exerted by its
`environment', i.e. the set of positions and velocities of all the
other surrounding particles. In the present framework we require
the functions $\mathbf{\Phi}_{t}^{i}$ to fulfill the following generally
admitted conditions: 
\begin{description}
\item [{{{\emph{a}.}}}] Two particles will interact provided they are
inside a neighborhood of fixed radius $R$. 
\item [{{{\emph{b}.}}}] The intensity of the interaction between two
neighboring particles should decay as a function of their interdistance. 
\item [{{{\emph{c}.}}}] In absence of any neighbor, particles should
follow an unbiased, completely erratic trajectory. 
\item [{{{\emph{d}.}}}] Frontal collisions between pairs of particles
should be hindered. 
\item [{{{\emph{e}.}}}] The interactions within a group of particles
should lead to the emergence of coherent patterns of collective motion. 
\item [{{{\emph{f}.}}}] The cluster formations made by particles should
maintain a certain degree of cohesion. 
\end{description}
\begin{figure}[h]

{\centering \includegraphics[width=0.5\textwidth,height=0.3\textheight]{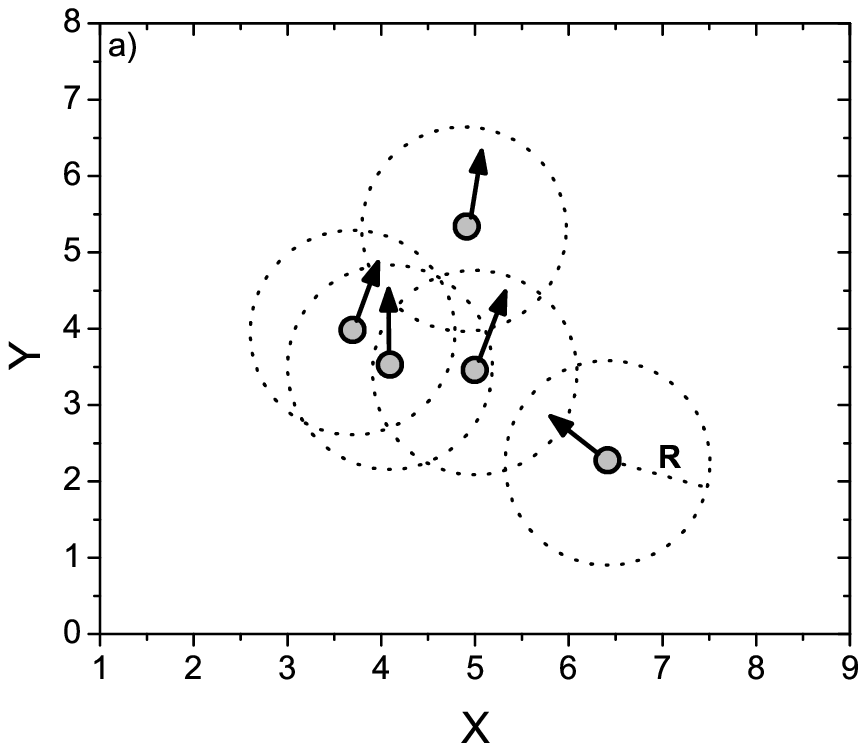}
\includegraphics[width=0.5\textwidth,height=0.3\textheight]{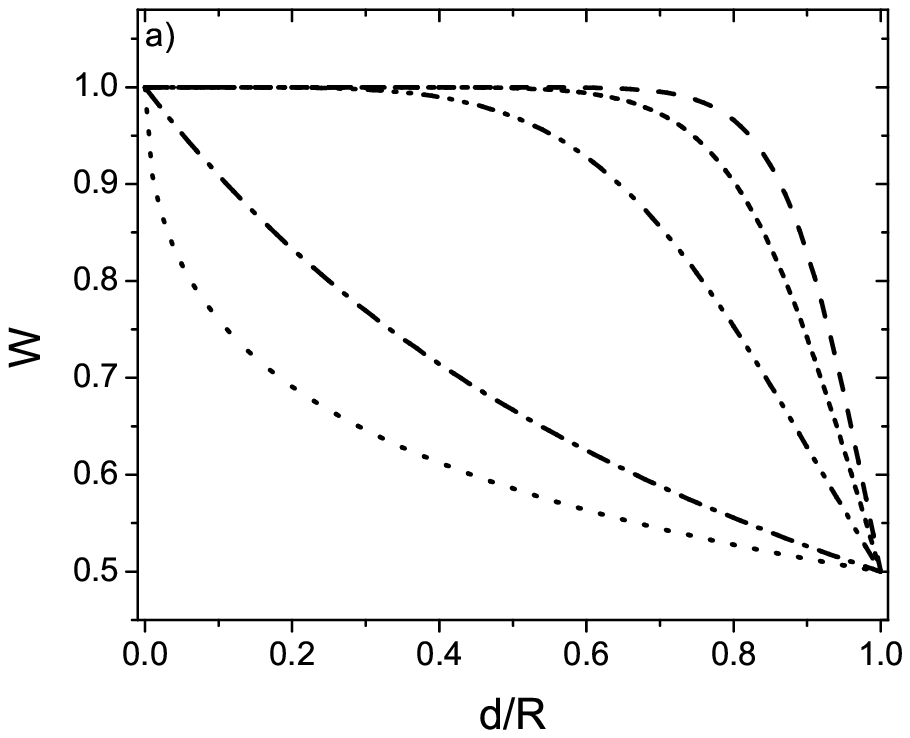}
} 

\caption{a) Particles interact via short range, local interactions extending
over a neighborhood of radius $R$. b) The graph of the weight function
\eqref{eq:W} where $d=\Vert\vec{r}_{t}^{i}-\vec{r}_{t}^{j}\Vert$
denotes the interparticle distance.}

\label{fig:RandW} 
\end{figure}

The coalescence of coherence and novelty observed in real collectives
of motiles suggests to consider an heuristic function $\Phi_{t}^{i}$
being able to display behaviors ranging from order to fully developed
chaos. As well known, the logistic map $x_{t+1}=\mu x_{t}(1-x_{t})$
with $x\in[0,1]$ and $0<\mu\leq4$ \cite{Nicolis99}, is a minimal
representative model for the class of discrete dynamical systems unfolding
the route to chaos via a period-doubling bifurcation cascade. Hence,
we propose the following function \begin{equation}
\mathbf{\Phi}_{t}^{i}=\Delta_{0}-2\cdot\Delta_{0}\cdot\phi_{t}^{i}\cdot\left(1-\left(\frac{\theta_{t}^{i}}{\Delta_{0}}\right)^{2}\right)\label{eq:logiFI}\end{equation}
 which is obtained by introducing the change of variable \begin{equation}
x_{t}\rightarrow\frac{1}{2}\left(1-\frac{\theta_{t}}{\Delta_{0}}\right)\mathrm{\; and\;\mu\rightarrow4\phi}\label{eq:Logi-transform}\end{equation}
 in the logistic map. Here, the functions $\phi_{t}^{i}\Bigr(\left\{ \vec{r}_{t}^{j},\vec{v}_{t}^{j}\right\} _{j\neq i}\Bigl)$
in \eqref{eq:logiFI} should embody the coupling between particles
in such a way that conditions \textbf{\textit{a}} - \textbf{\textit{f}}
are satisfied. As it is shown throughout the forthcoming sections,
the latter is achieved by the following function \begin{equation}
\phi_{t}^{i}=\begin{cases}
1-\frac{1}{2sn_{i}}\;\underset{j\in D_{i}(R)}{\sum}\Vert\vec{v}_{t}^{i}-\vec{v}_{t}^{j}\Vert w_{i,j}\; & \textrm{if }n_{i}>0\\
1 & \textrm{if }n_{i}=0\end{cases},\label{eq:paraFi}\end{equation}
 where $n_{i}$ is the number of neighboring particles counted within
a neighborhood $D_{i}(R)$ of fixed radius $R$ around each particle
$i$. Such neighborhood guarantees the validity of condition \textbf{\emph{a}}.
The interaction of neighboring particles $i$ and $j$ is weighted
by a suitably chosen function $w_{i,j}$ which aims at satisfying
condition \textbf{\emph{b}}. In particular, we shall assume that $w_{i,j}$
is given by \begin{equation}
w_{i,j}=\frac{R^K}{R^K+\left(\Vert\vec{r}_{t}^{i}-\vec{r}_{t}^{j}\Vert\right)^{K}}.\label{eq:W}\end{equation}
 Here, the parameter $K$ controls the degree of dependence of the
coupling on the interparticle distance. Notice that function \eqref{eq:W}
approaches a step function as $K$ increases (see Fig. \ref{fig:RandW}(a)).
As it turns out, for $K\gg1$ the effect of different interparticle
distances practically disappears, as the contributions of all interparticle
couplings, within the neighborhood, tend to be equally weighted.

\section{Local Dynamics \label{sec:LocalandMicro}}

We proceed now to the more `microscopic' local level in order to
elucidate the underlying dynamics of particle-particle interactions.
It is instructive to consider one particle in isolation and subsequently
a single pair of particles and their interaction. In a sense this
is analogous to the statistical mechanical treatment of `dilute gas'
where the particles' collisions, being rare, are described very accurately
by their binary collisions. Since, at very low densities, binary interactions
are the most dominant contributions in the present
model, we shall consider such a scenario in this
section.

\subsection{The one particle case}

\label{sec:The one particle case}

\begin{figure}[h]
{\centering \includegraphics[width=1\textwidth,height=0.4\textheight]{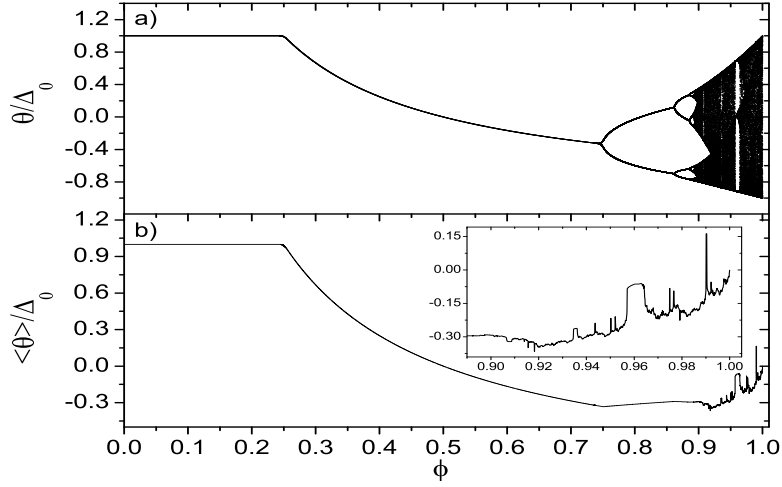}
} 

\caption{a) Bifurcation diagram of $\theta_{t}/\Delta_{0}$ as a function of
the parameter $\phi$. b) Similar plot for the mean phase $\langle\theta_{t}\rangle/\Delta_{0}$
(see eq. \eqref{eq:av-theta}). The inset of panel (b) provides a zoom-in
for $\langle\theta_{t}\rangle/\Delta_{0}$ in the $\phi$-interval
$(0.89,1]$.}

\label{fig:Bifurcationes} 
\end{figure}

Let us consider first a single particle with its parameter $\phi_{t}^{1}=\phi$
being a constant. In the present case, the whole information about
the possible types of particle trajectories is contained in the bifurcation
scenario for the variable $\theta_{t}$ as a function of $\phi$.
Such bifurcation diagram is qualitatively the same as in the logistic
map, as it is carried through by the transformations \eqref{eq:Logi-transform}
and $\phi=\mu/4$. As shown in Fig. \ref{fig:Bifurcationes}a), the
bifurcation diagram of the rotation phase depicts fixed points in
the interval $[0,\frac{3}{4}]$, the emergence of multiple period
orbits {[}MPO{]} in the interval $(\frac{3}{4},F_{p}]$ (where $F_{p}\approx0.8925$
corresponds to the well-known Feigenbaum accumulation point \cite{Nicolis99}),
onset of chaos at $F_{p}$, coexistence of weak chaos and periodic
motion in $(F_{p},1)$ and finally, fully developed chaos at $\phi=1$.
It is emphasized here that for $0<\phi<\frac{1}{4}$ the rotation
phase attains a maximum value corresponding to the trivial stationary
solution $\theta=\Delta_{0}$ of eqs. \eqref{eq:bigFI} - \eqref{eq:logiFI}.
For $\frac{1}{4}<\phi<\frac{3}{4}$ the rotation phase is given by
$\theta=\frac{1-2\phi}{2\phi}\Delta_{0}$. The particle thereby follows closed
trajectories with a radius of curvature growing as $\phi\rightarrow\frac{1}{2}$.
At $\phi=\frac{1}{2}$ a change in the sign of the rotation occurs,
from anticlockwise for $0\leq\phi<\frac{1}{2}$ to clockwise for $\frac{1}{2}<\phi\leq\frac{3}{4}$.
In order to characterize the trajectories of the particles in the
interval $[\frac{3}{4},1]$, it is worth considering the mean rotation
phase

\begin{equation}
\left\langle \theta\right\rangle =\frac{1}{m}\sum_{k=1}^{m}\theta^{(k)}\label{eq:av-theta}\end{equation}
 where \textit{m} denotes the period of a certain orbit and $\theta^{(k)}$
denotes its $k^{\mbox{\mbox{\textit{th}}}}$ element. In the $\phi$-interval
$(\frac{3}{4},0.862]$, which corresponds to orbits of period 2, the
mean rotation phase reads $\left\langle \theta\right\rangle =\frac{\Delta_{0}}{4\phi}$.
Providing an analytic expression for $\left\langle \theta\right\rangle $
in the case of orbits of period higher than 2 is a cumbersome task
and therefore, for $0.862<\phi\leq1$, we calculate numerically the
values of $\left\langle \theta\right\rangle $. These results are
presented in panel (b) of Fig. \ref{fig:Bifurcationes}. For $\phi$
values in the interval $(0.862,F_{p})$, the trajectories are closed
and clockwise. For $\phi$ in the interval $(F_{p},1)$, the mean
phase $\left\langle \theta\right\rangle $ displays highly irregular
behavior, where the alternation of windows of weak chaos and periodic
behavior gives rise to a fine structure profile. In the case of weak
chaos, the particle performs a biased chaotic walk leading to quasi-circular
trajectories. All along the interval $(F_{P},1)$, the particle motion
is clockwise on the average. The peaks observed in Fig. \ref{fig:Bifurcationes}(b),
where $\left\langle \theta\right\rangle $ attains higher values, are
indicative of the presence of windows of periodicity. Finally, at
the limiting case of fully developed chaos (i.e. $\phi=1$), the rotation
phase takes values distributed within the interval $[-\Delta_{0},\Delta_{0}]$,
according to the invariant density \[
\varrho_{s}(\theta,\phi=1)=\frac{1}{\pi\sqrt{\Delta_{0}^{2}-\theta^{2}}}\]
 of map \eqref{eq:bigFI} - \eqref{eq:logiFI}. The even character
of $\varrho_{s}$ entails that the particle describes a strongly
chaotic walk with $\left\langle \theta\right\rangle =0$, as it is confirmed by
 Fig. \ref{fig:Bifurcationes}(b). This way condition \textbf{\emph{c}}
above is satisfied since, according to \eqref{eq:paraFi}, all isolated
individuals evolve with $\phi_{t}^{i}=1$. This analysis has been
carried out for a single particle with an arbitrary fixed value of
$\phi$. Yet, the same categorization readily applies to the case
of many interacting particles, as long as their coupling parameters
$\phi_{t}^{i}$ converge towards a quasi-stationary value. The main
results obtained from the one particle case are summarized in Table
I.

\begin{longtable}{>{\centering}m{0.0033\textwidth}c|>{\centering}m{3cm}|}
\caption{Classification of the asymptotic behavior of $\theta$ and of the
particle trajectories, for characteristic domains of the parameter
$\phi$.}
\endfirsthead
\hline 
\multicolumn{1}{|c||}{\noun{interval: parameter range}} & \multicolumn{1}{c||}{\noun{asymptotic behavior of $\theta$}} & \noun{type of trajectory}\tabularnewline
\hline
\hline 
\multicolumn{1}{|l||}{A: $\phi\in[0,\frac{3}{4})$ } & \multicolumn{1}{c||}{stationary state} & closed, anticlockwise trajectories for $0<\phi<1/2$; straight trajectories
at $\phi=1/2$; clockwise, closed trajectories for $1/2<\phi<3/4$\tabularnewline
\hline 
\multicolumn{1}{|l||}{B: $\phi\in[\frac{3}{4},F_{p})$} & \multicolumn{1}{c||}{multiple period orbits} & closed, clockwise trajectories\tabularnewline
\hline 
\multicolumn{1}{|l||}{C: $\phi\in[F_{p},1)$} & \multicolumn{1}{c||}{weak chaos} & biased chaotic trajectories (quasi-closed, clockwise trajectories)\tabularnewline
\hline 
\multicolumn{1}{|l||}{D: $\phi=1$} & \multicolumn{1}{c||}{fully developed chaos} & unbiased chaotic trajectories \tabularnewline
\hline
\end{longtable}

\subsection{The two particle case: interparticle synchronization \label{sec:The two particle}}

The case of two particles is useful to consider in order to illustrate
some of the control features exercised by the coupling function \eqref{eq:paraFi}
which, in this particular case, is the same for both particles, $\phi_{t}^{(1)}=\phi_{t}^{(2)}\equiv\phi_{t}^{(1,2)}$.
Unless stated otherwise, the set of parameters we use throughout the
rest of the paper are: $s=1$, $\Delta t=1$, $R=500$, $\Delta_{0}=\frac{1}{40}\pi$
and $K=1$.

\begin{figure}[h]
 {\centering \includegraphics[width=1\textwidth,height=0.4\textheight]{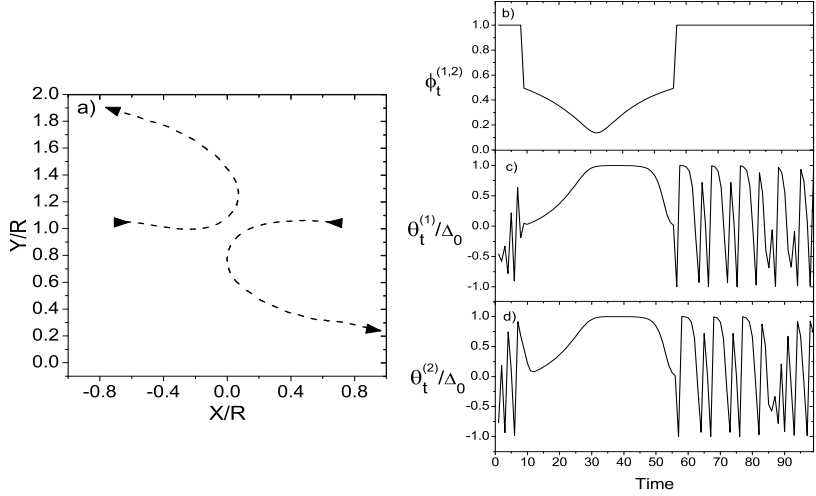}
} 

\caption{a) Trajectories of two particles initially approaching each other
frontally. b) Plot of $\phi_{t}^{(1,2)}$ versus time $t$, corresponding
to particles of panel a). c) Similar plot for $\theta_{t}^{(1)}/\Delta_{0}$
and $\theta_{t}^{(2)}/\Delta_{0}$. The parameters used are $s=1$,
$\Delta t=1$, $R=500$, $\Delta_{0}=\frac{1}{40}\pi$ and $K=1$.}

\label{Flo:fig2} 
\end{figure}

First, it is straightforward to show how condition \textbf{\textit{d}}
is satisfied. Figure \ref{Flo:fig2} depicts the plot of the trajectories
of two particles initially being out of the interaction zone and approaching
each other along the same axis. Under such conditions, when they both
enter the interaction zone, the difference of velocities approaches
its maximum value $\Vert\vec{v}_{t}^{(1)}-\vec{v}_{t}^{(2)}\Vert\approxeq2s$
and $w_{i,j}=\frac{1}{2}$. Therefore, as soon as both particles start
interacting, the parameter $\phi_{t}^{(1,2)}$ drops sharply from
unity to $\frac{1}{2}$, as it is shown in Fig. \ref{Flo:fig2}(b).
According to Fig. \ref{fig:Bifurcationes}(a), a value of $\phi_{t}^{(1,2)}<\frac{3}{4}$
entails a single fixed point for both phases $\theta_{t}^{(1)}$ and
$\theta_{t}^{(2)}$. As a result, the difference $|\theta_{t}^{(1)}-\theta_{t}^{(2)}|$
will tend to vanish after a brief transient. The latter can be realized
by comparing Figs. \ref{Flo:fig2}(b,c,d).

Moreover, since for $\Vert\vec{r}_{t}^{(1)}-\vec{r}_{t}^{(2)}\Vert\rightarrow0$
the weight $w_{(1,2)}\rightarrow1$, a further decrease in $\phi_{t}^{(1,2)}$
occurs as both particles get closer (see the decrease towards a minimum
value in the plot of Fig. \ref{Flo:fig2}(b)). As it was discussed
in Subsection \ref{sec:The one particle case}, a low value of $\phi_{t}^{(1,2)}<\frac{1}{3}$
leads both phases to reach their maximum value $\Delta_{0}$ (see
Figs. \ref{fig:Bifurcationes}a) and b)). This fact is demonstrated
by the onset of the plateau appearing in Figs. \ref{Flo:fig2}(c,d). Once the phases of the particles reach their maximum value, simultaneously
with their minimum phase difference, they follow their course away
from each other. As it turns out, condition \textbf{\emph{d}} is satisfied
as both particles turn away sharply instead of colliding frontally
(see Fig. \ref{Flo:fig2}(a)). Eventually, both particles leave the
interaction zone and display again fully erratic trajectories. %
\begin{figure}[h]
 { \includegraphics[width=1\textwidth,height=0.4\textheight]{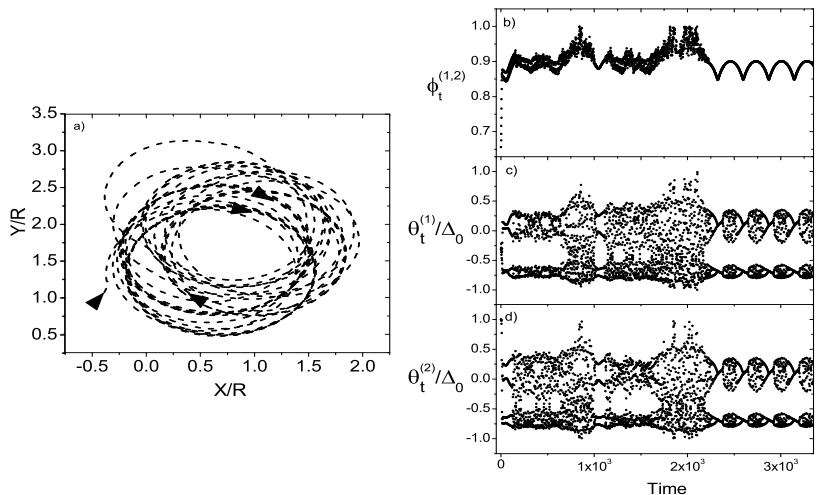}
} 

\caption{a) Trajectories of two particles located initially inside the interaction
zone and approaching each other with a moderate difference of initial
velocities. In panel b) we present the plot of $\phi_{t}^{(1,2)}$
versus time $t$ corresponding to particles of panel a). In panels
c) and d), we show similar plots for $\theta_{t}^{(1)}/\Delta_{0}$
and $\theta_{t}^{(2)}/\Delta_{0}$ respectively. The parameters we
used in these plots are the same as in Fig. \ref{Flo:fig2}.}

\label{Flo:fig5} 
\end{figure}

\begin{figure}[h]
{\centering \includegraphics[width=1\textwidth,height=0.4\textheight]{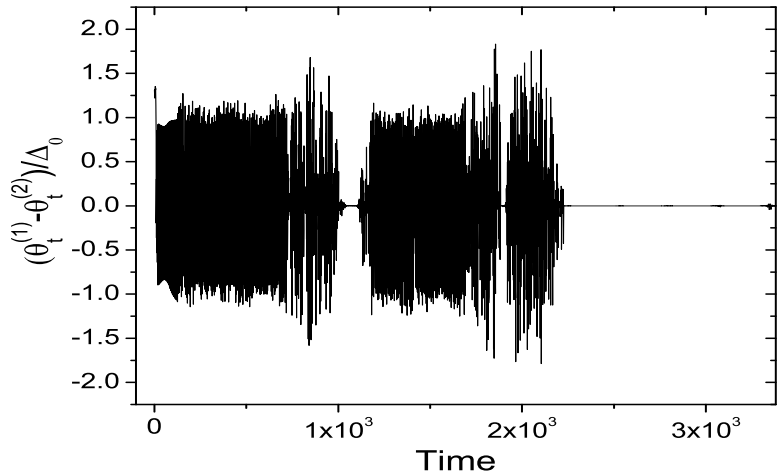}
} 

\caption{Corresponding to phases $\theta_{t}^{(1)}$ and $\theta_{t}^{(2)}$
of Fig. \ref{Flo:fig5}, the plot of the time evolution of the phase
difference $(\theta_{t}^{(1)}-\theta_{t}^{(2)})/\Delta_{0}$.}

\label{Flo:fig:6} 
\end{figure}

Let us consider now the case of two particles approaching each other
inside the interaction zone with a small difference in their initial
directions. Figure \ref{Flo:fig5}(a) shows the trajectories of two
particles in such a typical situation, while Fig. \ref{Flo:fig5}(b)
depicts the evolution of the coupling parameter $\phi_{t}^{(1,2)}$
and Figs. \ref{Flo:fig5}(c,d) show the evolution of the rotation
phases $\theta_{t}^{(1)}$ and $\theta_{t}^{(2)}$ of both particles.
Initially, a sharp decrease in the difference of directions $\Vert\vec{v}_{t}^{(1)}-\vec{v}_{t}^{(2)}\Vert$
occurs, which appears in the plot of Fig. \ref{Flo:fig5}(b) as an initial
abrupt increase of $\phi_{t}^{(1,2)}$. Since $\phi_{t}^{(1,2)}$
attains a value higher than $\frac{3}{4}$, the rotation phases of
both particles subsequently enter into the MPO and chaos regimes.
The ongoing evolution process is driven by two counteracting tendencies:
The first one is a trend towards strong synchronization if $\phi_{t}^{(1,2)}$
decreases below $F_{P}$. The second one is a trend to spread away
if an alignment occurs, i.e. if $\phi_{t}^{(1,2)}\rightarrow1$. The
presence of these two trends is revealed when comparing panels (a),
(b) and (c) of Fig. \ref{Flo:fig5}: the values of $\theta_{t}^{(1)}$
and $\theta_{t}^{(2)}$ tend to further localize when $\phi_{t}^{(1,2)}$
decreases, while bursts of chaos appear whenever $\phi_{t}^{(1,2)}\approx1$
(corresponding to an alignment). As it is depicted in Figs. \ref{Flo:fig5}(a-c), after a transient self-organization process, the evolution
of both particles undergoes a transition towards a highly coherent
dynamic regime. The initially irregular behavior observed in the course
of the evolution of $\phi_{t}^{(1,2)}$ suddenly disappears and, instead,
oscillations emerge within the $\phi$-interval $(0.85,0.9)$. Such
oscillations of $\phi_{t}^{(1,2)}$ induce on both phases $\theta_{t}^{(1)}$
and $\theta_{t}^{(2)}$ a regular alternation between windows of high
and low period orbits (see Figs. \ref{Flo:fig5} (c,d)). Once such
a new regime is attained, both particles tend to move in perfect synchrony
and their phase difference $\vert\theta_{t}^{(1)}-\theta_{t}^{(2)}\vert$
equals zero, as it is readily shown in Fig. \ref{Flo:fig:6}. As it
turns out, for the case of two particles, it is clear that at longer
term the interplay between both tendencies (ordering and chaotic phase
spreading) constitute a `control mechanism' that underlies the fulfillment
of conditions \textbf{\emph{e}} and \textbf{\emph{f}}. Using the set
of parameters as in Fig. \ref{Flo:fig2} for different settings of
initial conditions, we found that phase synchronization, as it is
observed in Fig. \ref{Flo:fig:6}, occurs provided the initial conditions
are such that both particles never leave the interaction zone. In
Section \ref{sec:The-Emergent-Global}, we show that phase synchronization
is robust in the case of many particles as well.

\section{Emergent global dynamics and patterns of motion \label{sec:The-Emergent-Global}}

Having seen the role of spontaneous synchronization in binary interactions
we can now consider the full collective motion problem. As we shall
demonstrate further in this section, synchronization is an underlying
mechanism for coherent and cohesive motion. Another feature of this
intrinsic and coherent steering, which keeps a large fraction of the group together, is
that the role of boundary conditions becomes really secondary. Furthermore, we emphasize the fact that the aggregates of motiles keep 
its coherent motion even in absence of boundaries.

So, let us address now the case of $N$ particles allowed to move on the
infinite plane. We consider, thereof, that the initial positions and
velocities of the particles are uniformly randomly distributed within
a square of side $R_{0}$. Similarly, the rotation phases are distributed
in the interval $[-\Delta_{0},\Delta_{0}]$.

In order to capture the spatio-temporal features of the system, such
as persistent and/or flickering clustering as well as patterns of
cohesion around temporary foci, it is of interest to introduce an
index that takes into account both the spatial relations between neighboring
particles and the degree of velocity alignments. To this aim, it is
instrumental to visualize the system as an undirected graph, where
every particle is its node and the edges of the graph are established according
to their interparticle distance and to the degree of alignment between
particle velocities. 

In particular, we consider a pair of nodes as connected if, in addition
of being neighbors, the inner angle between their velocities is less
than $\Delta_{0}$. Under these assumptions, the time dependent adjacency
matrix $\boldsymbol{M}_{t}$ associated to such a graph reads \begin{equation}
[M_{t}]_{i,j}=\begin{cases}
1 & \mbox{if }\Vert\vec{r}_{t}^{(i)}-\vec{r}_{t}^{(j)}\Vert\leq R\mbox{ and }\Vert\vec{v}_{t}^{i}-\vec{v}_{t}^{j}\Vert\leq s\Delta_{0}\\
0\mbox{ } & \mbox{otherwise}\end{cases}\end{equation}
\label{eq:M}In terms of this matrix, one can quantify the degree of collective
alignment resulting from interparticle interactions by means of the following index 
\begin{equation}
\alpha_{t}=\frac{\Gamma_{t}}{\Gamma_{M}},\end{equation}
\label{eq:FC}where $\Gamma_{M}=\tfrac{1}{2}N(N-1)$ is the maximum
number of possible connections and $\Gamma_{t}=\frac{1}{2}Tr\left[M_{t}^{2}\right]$
(with $Tr$ denoting the matrix trace and $M_{t}^{2}$ the square
of the adjacency matrix) gives the number of actual
connections at time $t$. Hereafter we refer to index $\alpha_t$ as the Alignment Clustering Index (ACI) at time $t$. Clearly, ACI is an indicator of the degree
of coherence of motion of the group of particles. Furthermore, it
provides information on the degree of clustering of particles in space
and thus it constitutes an indicator of the degree of group cohesion.
For instance, strong oscillations in the evolution of the ACI indicate
a poorer group cohesion, since its fluctuations reveal that a great
amount of particles are merging in as well as escaping from cluster
formations.

\subsection{Typical patterns and regimes of motion}

\label{sec:Typical patterns and regimes of motion}

In order to demonstrate some typical patterns of characteristic behavior
of our model, let us consider a collective of particles initially
distributed in a squared region of side $R_{0}=\dfrac{1}{3}R$. Unless
explicitly mentioned, we shall consider the number of particles $N=300$
and the rest of the parameters to be the same as in Fig. \ref{Flo:fig2}.
\begin{figure}
{\centering \includegraphics[width=1\textwidth,height=0.35\textheight]{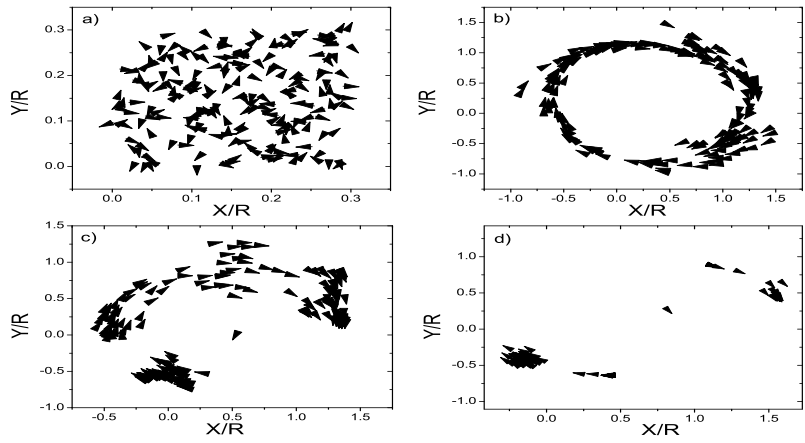}
} 

\caption{For a system of $N=300$ particles, in absence of boundaries and initially uniformly randomly settled
within a squared region of side $R_{0}=\frac{R}{3}$, plots of the
positions and velocities at a) $t=0$, b) $t=2\times10^{3}$, c) $t=6\times10^{3}$
and d) $t=1.5\times10^{4}$. The parameters are the same as in Fig.
\ref{Flo:fig2}.}

\label{Flo:fig7} 
\end{figure}

\begin{figure}[h]
{\centering \includegraphics[width=1\textwidth,height=0.4\textheight]{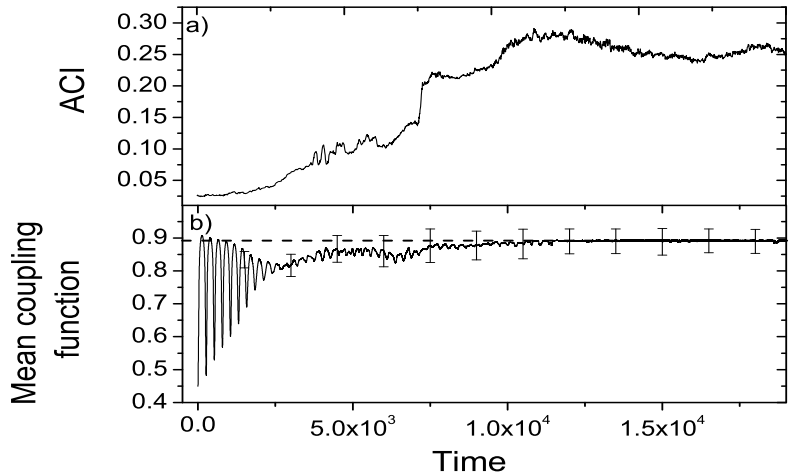}
} 

\caption{Corresponding to the realization of Fig. \ref{Flo:fig7}, the evolution
of the ACI is shown in panel a). In panel b), the mean value of the
coupling functions $\phi_{t}^{i}$ of all particles (where bars provide
the corresponding standard deviation) is presented. The dashed line
of panel a) corresponds to the value of the Feigenbaum point.}

\label{Flo:fig8} 
\end{figure}

Figure \ref{Flo:fig7} shows the plot of the positions and velocities
of the set of particles at different times in a single realization.
Corresponding to the system of Fig. \ref{Flo:fig7}, panel (a) of Fig.
\ref{Flo:fig8} shows the evolution of the ACI and panel (b) the evolution
of the first and second moments of the coupling functions \eqref{eq:paraFi}
of all particles. One can distinguish between three qualitatively
different stages of collective motion, which can be characterized
by means of the ACI behavior:
\begin{enumerate}
\item {Starting from a disordered state, Fig. \ref{Flo:fig7}(a), the system
undergoes a strong self-organization process, from $t=0$ up to $t\thickapprox2\times10^{3}$. This is characterized
by low ACI values (see Fig. \ref{Flo:fig8}(a)), and by large oscillations at the level
of the coupling functions, Fig \ref{Flo:fig8}(b). At the end of this
stage the collective motion becomes further organized and the group
of particles shapes a circular rotational pattern (see Fig. \ref{Flo:fig7}(b)).}
\item {The second stage, which extends approximately from $t\thickapprox2\times10^{3}$
up to $t\thickapprox1\times10^{4}$, is mainly characterized by a
marked increase of the ACI (see Fig. \ref{Flo:fig8}(a)). The coupling
functions tend to stabilize around a fixed value, as exhibited by
the decrease in the amplitude of oscillations of the mean coupling
(see Fig. \ref{Flo:fig8}(b)). Along this stage, an increasing number
of groups of highly aligned particles are formed (see Fig. \ref{Flo:fig7}(c)), thereby fulfilling condition \textbf{\emph{e}}.}
\item {At the final stage, the system attains a quasi-stationary regime,
at $t\thickapprox1\times10^{4}$, along which the ACI stabilizes and
the coupling functions remain nearly constant (see Figs. \ref{Flo:fig8}(a,b)).
In such a regime, the standard deviation of the average also tends
to stabilize around a rather low value, thereby indicating an enhancement
in the coherence of the behavior of the particles (see Figs. \ref{Flo:fig8}(c)
and \ref{Flo:fig7}(d)). Clearly, group cohesion is evidenced by the
fact that high values of the ACI are sustained for times as long as
$t=1\times10^{4}$ and beyond. As it turns out, condition \textbf{\emph{f}}
is also satisfied. It is a remarkable fact that the self-organization
process leading to such state of higher coherence drives also the
average of the coupling functions towards the region around the Feigenbaum
point (which is represented by a dashed line in Fig. \ref{Flo:fig8}(b)).
In other words, a higher degree of coherence is achieved by a process
that exhibits a combination of weak chaos and further ordered behavior.
This question will be discussed later in Subsection \ref{sec:Group cohesion},
where we shall address the cohesion features of the presented system.} 
\end{enumerate}

\subsection{Synchronization processes in the collective}

\label{sec:Synchronization processes} In Subsection \ref{sec:The two particle}
we have shown that the emergence of coherent behavior in the case
of two particles arises from synchronization at the level of their
rotation phases. Thus, it is natural to inquire on the degree of phase
synchronization underlying the coherent aspect of the dynamics of
the $N$-particle system as it is exhibited in the plots of Figs.
\ref{Flo:fig7} and \ref{Flo:fig8}. In order to assess interparticle
phase synchronization in a quantitative manner, we shall consider
the ensemble of the distinct interparticle phase differences $\Delta\theta_{ij}=\theta_{i}-\theta_{j}$
with $i\neq j$. We shall, thus, monitor the evolution
of the system in terms of its synchronized population fraction. Such
task is achieved by quantifying the percentage $P_{t}(\Delta\theta)$
of pairs of particles with a given phase difference $\Delta\theta$
at time $t$. Thereby, assessing the synchronization process amounts to follow the evolution of $P_{t}(\Delta\theta)$
on the $\Delta\theta$-interval $[-2\Delta_{0},2\Delta_{0}]$ starting
from an initial distribution $P_{0}$. Clearly, full synchronization
will be present whenever $P_{0}$ converges asymptotically towards
a delta distribution at $\Delta\theta=0$. %
\begin{figure}[h]
{\centering \includegraphics[width=1\textwidth,height=0.4\textheight]{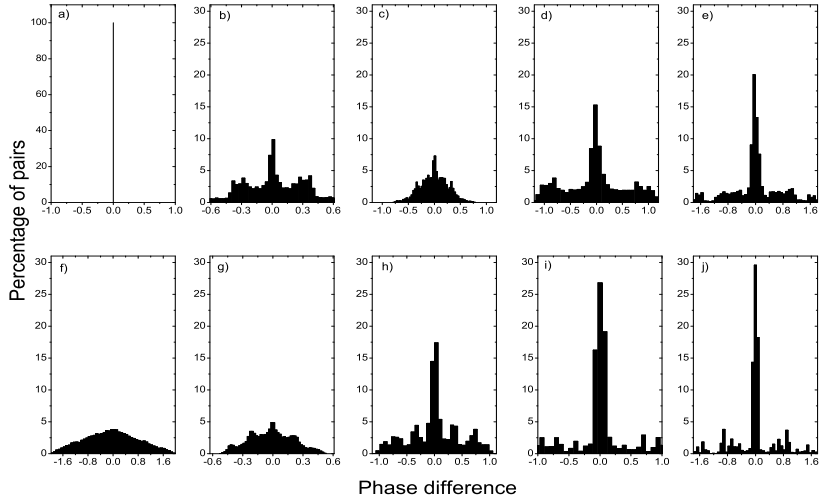}
} 

\caption{Plot of the evolution of two different initial distributions of phase
differences for $N=300$ particles. Upper panels: Having all particles
the same initial phase $\theta$, distribution of phase differences
at (a) $t=0$, (b) $t=2\times10^{3}$, (c) $t=1\times10^{4}$, (d) $t=4\times10^{4}$
and (e) $t=1\times10^{7}$. Lower panels: for particles with initial
phases uniformly randomly distributed, at (f) $t=0$, (g) $t=2\times10^{3}$,
(h) $t=1\times10^{4}$, (i) $t=4\times10^{4}$ and (j) $t=1\times10^{7}$.
The parameters used in this figure are the same as in Fig. \ref{Flo:fig2}.
Such numerical experiments supply ample evidence of the existence
of an attracting distribution.}

\label{Flo:syncN} 
\end{figure}

Figure \ref{Flo:syncN} exhibits the plot of the distribution $P_{t}(\Delta\theta)$
at different times (from $t=0$ up to $t=1\times10^{7}$), starting
from two different initial phase distributions: in upper panels (from
(a) to (e)), the evolution of an initial delta distribution. In lower
panels (from (f) to (j)), similar plots corresponding to an initial distribution
obtained by assigning uniformly distributed random values to the initial
set of phases $\{\theta_{t}^{i}\}$ (the triangular shape in the initial
distribution in Fig. \ref{Flo:syncN}(f) follows from the correlations
introduced by taking the differences between phases). After a very
short transient period, the initial correlations of phases corresponding
to both distributions are destroyed and new ones are built as a consequence
of the initial strong self-organization process described in Subsection
\ref{sec:Typical patterns and regimes of motion}. As it is shown
in the upper and lower panels of Fig. \ref{Flo:fig2}, phase synchronization
gradually emerges. At $t=1\times10^{4}$ the percentage of pairs with
phase difference $|\Delta\theta|<\Delta_{0}/10$ is $38\%$ for the
case of the initial delta distribution (Fig. \ref{Flo:syncN}(c)) and
$40\%$ for the case of the homogeneous initial distribution (Fig.
\ref{Flo:syncN}(h)). As the process evolves, the interparticle phase
synchronization is further enhanced. For instance, in the case of
the delta initial distribution, the percentage of pairs with $|\Delta\theta|<\Delta_{0}/100$
at $t=1\times10^{7}$ equals $36\%$ (Fig. \ref{Flo:syncN}(e)) and
it equals $60\%$ for the homogeneous one (Fig. \ref{Flo:syncN}(j)).
Since similar results are obtained for very different types of initial
distributions, one concludes that the synchronization phenomenon observed
in Fig. \ref{Flo:syncN} is robust for parameter values as in Fig.
\ref{Flo:fig2}. However, as it is shown in Subsection \ref{sec:Group cohesion},
similar self-organization processes are observed in a wide range of $K$-parameter
values.

\begin{figure}[h]
{\centering \includegraphics[width=1\textwidth,height=0.4\textheight]{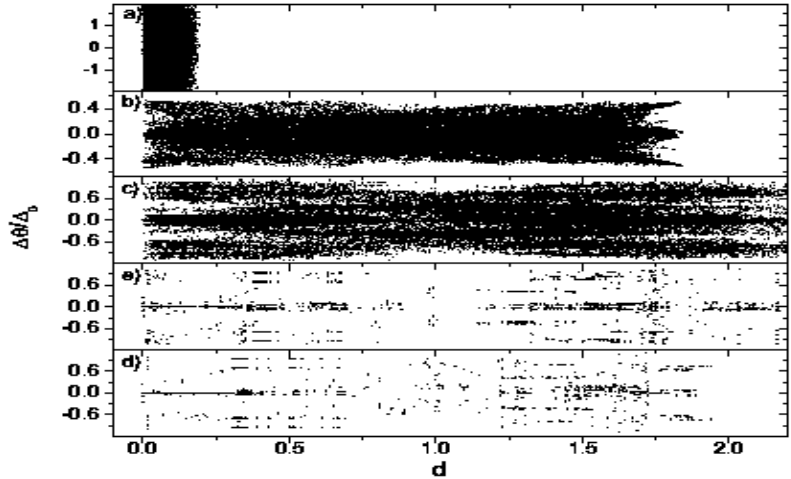}
} 

\caption{Corresponding to plots f) to j) of Fig. \ref{Flo:syncN}, plot of
phase differences ($\Delta\theta/\Delta_{0}$) versus interparticle
distances at a) $t=0$, b) $t=2\times10^{3}$, c) $t=1\times10^{4}$,
d) $t=4\times10^{4}$ and e) $t=1\times10^{7}$.}

\label{Flo:Phaseclust} 
\end{figure}

Regarding the relation between the emergence of synchronization and
the gradual formation of clusters of aligned particles in space (see
Figs. \ref{Flo:fig7}(c,d)), it is interesting to inspect the evolution
of the relation between pair synchronization and interparticle distance.
Figure \ref{Flo:Phaseclust} depicts the plots of phase difference
\emph{vs} interparticle distances for the particle pairs accounted
for in the histograms of panels (f) - (j) of Fig. \ref{Flo:syncN}.
Initially, all pairs are close to each other, having phase differences
randomly distributed in the interval $[-2\Delta_{0},2\Delta_{0}]$
(Fig. \ref{Flo:Phaseclust}(a)). At $t=2\times10^{3}$, particles are
further spread away and higher density `clouds' appear within a narrower
$\Delta\theta$-interval (Fig. \ref{Flo:Phaseclust}(b)). The evolution
towards a higher degree of organization leads to cluster formation
in both space and phase-difference. As it is displayed in Fig. \ref{Flo:Phaseclust}(c),
at $t=1\times10^{4}$ most of the pair particles are distributed in
two large clusters. Strong synchronization ($\Delta\theta\approx0$)
occurs between particles within the same cluster, as well as between
particles belonging to different clusters. Also, many pairs inside
and between clusters exhibit phase differences around specific values
which correspond to the secondary peaks in the distribution of Fig.
\ref{Flo:syncN}(h). For a longer time, i.e. $t=4\times10^{4}$, the
main clusters split into smaller lumps within which particles are
further synchronized with each other (Fig. \ref{Flo:Phaseclust}(e)).
At $t=1\times10^{7}$, a single compact cluster of fully synchronized
pairs ($\Delta\theta=0$) still remains. As it turns out, the presence
of strong synchronization at long term entails also a high degree
of group cohesion.

In comparison with other models of collective behavior, such as Vicsek's,
one naturally expects the degree of group cohesion to decrease along
with the initial density of particles. This question, as well as the
dependence of the cohesion on the parameter $K$ in the weight function
\eqref{eq:W}, are addressed in the following Subsection.

\subsection{Group cohesion: the role of density and interparticle distances}

\label{sec:Group cohesion} 
\begin{figure}
{\centering \includegraphics[width=1\textwidth,height=0.3\textheight]{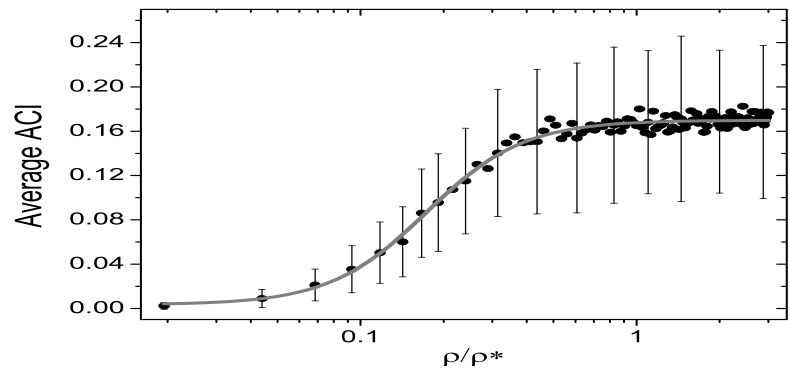}
} 

\caption{Plot of the ensemble average of the ACI, $\left\langle \alpha_{t}(\rho/\rho^{*})\right\rangle $,
of $N=100$ particles at time $t=4\times10^{4}$ (dots) for different
values of the logarithm of the ratio $\rho/\rho^{\star}$. Here the
average was taken over an ensemble of $200$ uniformly random initial
conditions. The rest of the parameters are the same as in Fig. \ref{Flo:fig2}.
Bars in the plot indicate the standard deviation. The curve described
by these points is best fitted by a logistic function in the form
of Eq. \eqref{eq:ACI_vs_density}.}

\label{Flo:figAverageFCvsDensity} 
\end{figure}

In the context of animal societies, sharp
transitions of collective behavior have been reported to occur at
specific values of the population density (see for instance \cite{SFaEWaTB01,Aldana07,DeneubourgBook,Dossetti09,GGaHC04,MNaIDaTV07,OJOaMRE99,WLaHZaMZCaTZ08}).
As it has been already mentioned in the introduction, models belonging
to the same class as Vicsek's exhibit an order-disordered phase
transition at a critical value of the particle density. In the model
introduced here, the radius of interaction $R$ imposes a characteristic
density $\rho^{\star}=N/R^{2}$. Thus, it is natural to study the
degree of coherence and cohesion in a group of particles for density
values above and below $\rho^{\star}$. Figure \ref{Flo:figAverageFCvsDensity}
depicts the plot of the ensemble average of the ACI for a group of $N=100$
particles at $t=4\times10^{4}$ and for different values of the ratio
$\rho/\rho^{\star}$. This plot exhibits a steep
monotonic ascent of the averaged ACI occurring for values $\rho/\rho^{\star}\sim O(10^{-2})$,
followed by the onset of a plateau extending for values $\rho/\rho^{\star}>1$.
Such a behavior is best fitted by a logistic function of the form
\begin{equation}
\left\langle \alpha(\rho/\rho^{*})\right\rangle_{t=4\times10^{4}}=a+\cfrac{b}{c^{p}+(\rho/\rho^{*})^{p}}\label{eq:ACI_vs_density}\end{equation}
where $a=0.170$, $b_{0}=0.003$, $c=0.173$ , $b=(b_{0}-a)c^{p}$
and $p=2.460$ (as it is represented in Fig. \ref{Flo:figAverageFCvsDensity}
by a continuous gray line) with a correlation coefficient of about
$0.976$.

It must be noted here that the choice of the logarithmic scale for
the ratio $\rho/\rho^{\star}$ serves in elucidating the abrupt change
observed. The change of behavior exhibited in Fig. \ref{Flo:figAverageFCvsDensity},
as described by the logistic form of Eq. \eqref{eq:ACI_vs_density},
differs in nature from the phase transition reported for Vicsek's
class of models \cite{MNaIDaTV07,HCaFGaGGaFPaFR08}. In addition,
we stress the fact that the abrupt change in the degree of coherence
shown in Fig. \ref{Flo:figAverageFCvsDensity} arises naturally from
the interparticle interactions, through concerted self-adaptation
in the steering parameters of the particles.

\begin{figure}[h]
{\centering \includegraphics[width=1\textwidth,height=0.4\textheight]{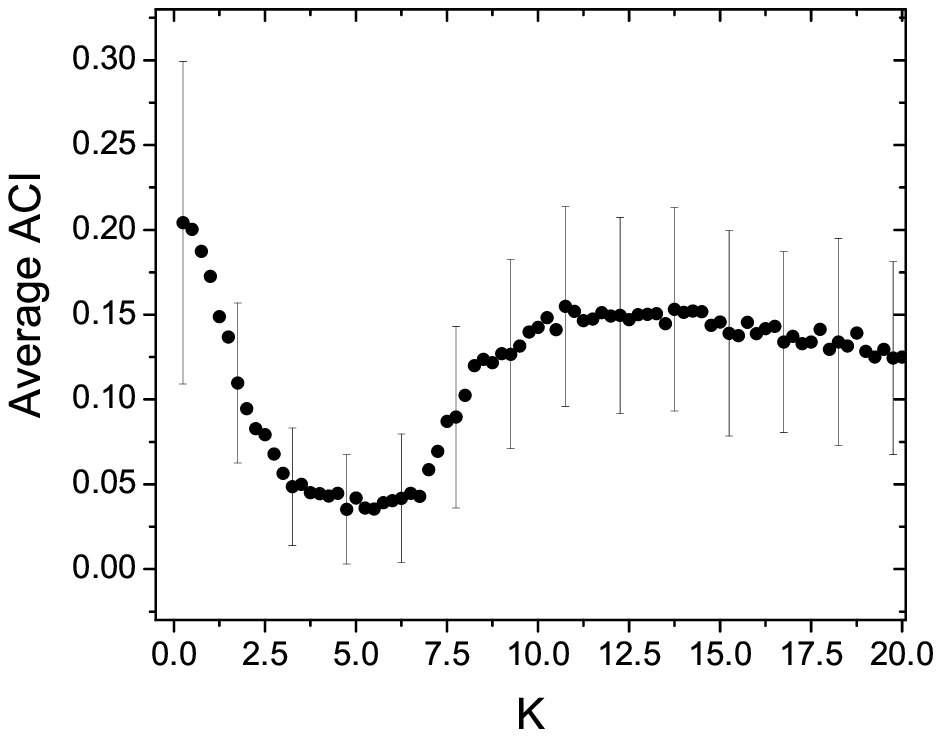}
} 

\caption{For a group of $N=100$ particles, plot of the ensemble average of
the ACI, at $t=4\times10^{4}$ for different values of the parameter
$K$. Similarly as in Fig. \ref{Flo:figAverageFCvsDensity}, the average
is computed over an ensemble of 200 uniformly randomly distributed
initial conditions where error bars indicate one standard deviation.}

\label{Flo:figAverageFC_vs_K} 
\end{figure}

On the other hand, the cohesion of the group is controlled additionally
by the parameter $K$ in the weight function \eqref{eq:W}. Figure
\ref{Flo:figAverageFC_vs_K} shows the plot of the ensemble average
of the ACI at $t=4\times10^{4}$ for different values of $K$. One
observes that stronger cohesion, as gauged by the ACI, is attained for small values of $K$. For $0<K\leq1.25$, the average
ACI takes values around $15-20\%$. A decrease in the averaged ACI
is observed in the interval $0<K\lesssim5$ which is followed by an
increase for $5<K\lesssim10$. For $K$ values greater than 10, the
averaged ACI gradually decreases. It is noticeable that the change
in the overall behavior of the ACI around $K=5$, from decrease to
increase, coincides with the change of sign in the curvature of the
graph of weight functions $W_{{i},{j}}$ of \eqref{eq:W} when varying
$K$. Since the form of the weight functions ultimately affects the
behavior of the coupling functions $\phi_{i}$ of \eqref{eq:paraFi},
it is worth considering the ensemble average of the histograms of
the coupling functions of all particles. In particular, we shall focus
on the main regions described in Table I, namely \textbf{A}: a steady
state, \textbf{B}: multi-periodicity, \textbf{C}: weak chaos and \textbf{D}:
fully developed chaos. 

\begin{figure}[h]
 {\centering \includegraphics[width=1\textwidth,height=0.4\textheight]{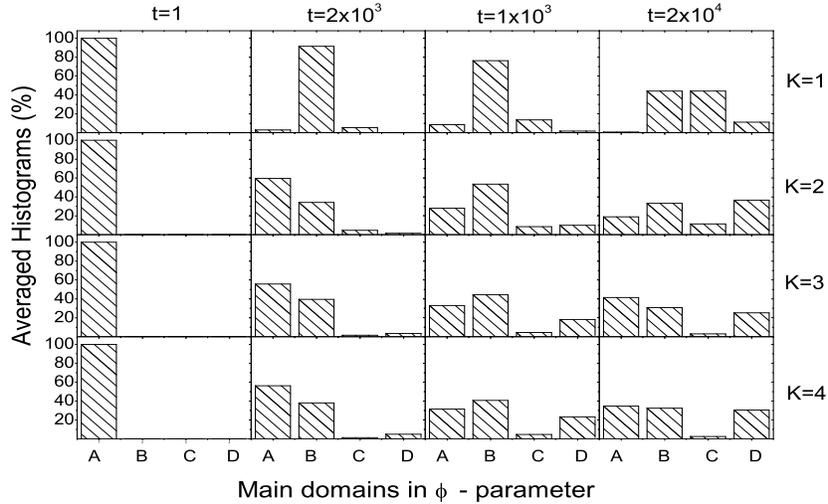}
} 

\caption{Plots of the ensemble average of histograms obtained from a group
of $N=100$ particles for the $\phi$-intervals of Table I which
correspond to \textbf{A}: a steady state, \textbf{B}: multi-periodicity,
\textbf{C}: weak chaos and \textbf{D}: fully developed chaos (the
percentage of particles in \textbf{D} indicates the proportion of
isolated particles). The average was taken over an ensemble of 200
initial conditions. Columns from left to right in this plot present
results obtained for $t=1$, $t=2\times10^{3}$, $t=1\times10^{4}$
and $t=2\times10^{4}$ while rows from top to bottom correspond to
results obtained for $K=1$, $K=5$, $K=10$ and $K=15$. The rest
of the parameters are the same as in Fig. \ref{Flo:fig2}.}

\label{Flo:figCat-Histograms-1} 
\end{figure}

Figure \ref{Flo:figCat-Histograms-1} displays the ensemble average
of the histograms obtained at different times and for different values
of $K$. Initially (first column of Fig. \ref{Flo:figCat-Histograms-1}),
regardless of the value of $K$, one observes that the coupling functions
take values in the region where a steady state exists. For $K=1$
(first row of Fig. \ref{Flo:figCat-Histograms-1}), the values of
the couplings gradually evolve towards the region of multi-periodicity
and beyond. At $t=4\times10^{4}$ one observes that the bulk of particles
is equally distributed between the multi-periodicity and weak chaos
regions. However, $10\%$ of the particles also appear in the region
of fully developed chaos, which reveals particle escapes away from
the center of mass. In the cases $K=5$, $K=10$ and $K=15$ (rows
3, 4 of Fig. \ref{Flo:figCat-Histograms-1}), at long times, more
than $20\%$ of the particles have escaped and those that remain confined
exhibit $\phi_{i}$-values mostly in the regions corresponding to
stationarity and multi-periodicity. Summarizing, in the case $K=1$,
a higher degree of coherence and group cohesion is attained through
a process encompassing balance between weak chaos and ordered behavior.
In the cases $K=5$, $K=10$ and $K=15$, where most of the interacting
particles display further ordered behavior, the degree of cohesion
is poorer. In particular, in the case $K=5$ almost $40\%$ of the
particles have escaped at $t=4\times10^{4}$ (as it is indicated by
an increased number of particles with $\phi_{j}\in D$). Consequently,
for $K=5$, the averaged ACI attains a very low value of approximately
0.05 (see Fig. \ref{Flo:figAverageFC_vs_K}). For $5<K<10$, a further
rigid steering motion arises, as an increased number of particles
having $\phi_{j}\in\mbox{A}$ tend to display short concentric trajectories.
Thereby, the number of escapes is smaller and the average ACI becomes
higher than in the case of $K=5$ (see Fig. \ref{Flo:figAverageFC_vs_K}).
The latter provides an insight into the origin of the minimum of the
averaged ACI observed around $K=5$.

\section{Conclusions \label{sec:Conclusions}}

In this work we have introduced a minimal model of motile particles
which takes into account an intrinsic steering mechanism. Its main
feature is that it exhibits the emergence of patterns of coherent
collective motion. Since each particle is endowed with an intrinsic
mechanism, it allows it to adjust its trajectory according to the
surrounding conditions. With this extra feature the model can be of
utility as an augmented inert particle model. The adaptation is achieved
by changes which are determined by a map of the logistic-map family,
which displays fully developed chaos in absence of interparticle interactions.
As it turns out, isolated particles behave as chaotic walkers. On
the other hand, particles within a neighborhood of fixed radius interact
with each other by establishing nonlinear couplings between their
corresponding logistic maps. The coupling between a particle and its
surrounding neighbors is embodied by a function that depends explicitly
on the positions and velocities of the set of involved particles.
The coupling functions play their role in self-adapting the controlling
parameters of the logistic map at the individual level of each motile
by `tuning in' the behavior of each particle. This behaviour can range
from single to multiple periodic and chaotic regimes. The explicit
form of the coupling functions is such that frontal collisions are
hindered and that neighboring particles undergo a self-organization
process leading to the emergence of coherent collective motion. The
whole process is characterized by an index, called ACI (and denoted
by $\alpha$ herein). This graph index, captures the essential features
of the system's time evolution, since it quantifies the degree of
cohesion and alignment of particles. It is shown that the self-organization
process of steering leads the collective of these motiles towards
a regime of coherence and group cohesion. The latter is shown to be essentially related 
to the phase synchronization that emerges between the inner steering mechanisms of the particles.

Also, the degree of group coherence is studied as a function of a
parameter $K$. This parameter sets the dependence of the interactions
on the interparticle distances. Such a study shows that higher levels
of coherence and group cohesion occur in cases where most of the particles
exhibit a balanced combination of ordered motion (multi-periodicity)
and weakly chaotic behavior.

Similarly, as in other reported cases, the system presented here exhibits
a drastic change in the degree of coherence as a function of the initial
particle density but in a more abrupt fashion. However, such result
does not require the use of periodic boundary conditions as it is
customary the case in other classes of models.

Further investigation of this model is required in order to develop
insights on key questions regarding the phenomenon of phase synchronization
observed, as well as, the dependence of the characteristic time of
synchronization on the model parameters and on how robustness of synchronization is affected by the density of particles.

On the other hand, since our system can be considered of the minimal
ones that includes a deterministic steering, it can be readily augmented
in order to account for more realistic situations. Depending on the
context of application one might easily consider further the inclusion
of particle accelerations, different types of long or short range
interactions for the interparticle attraction, inclusion of inhomogeneities
and ambient gradients effectuated at the level of the interactions
and finally the introduction of population variability such as groups
of leading and/or inert particles.

\section*{Acknowledgments}

The authors would like to thank G. Nicolis, J - L. Deneubourg and T.
Bountis for offering their encouragement, insightful comments and
most fruitful criticism during the preparation of the present paper.
We would, also, like to thank E. Toffin and M. Lefebre for motivating
this work and for suggesting relevant references. The work of A. G. C. R. was supported by the `Communaut\'{e} Francaise de Belgique' (contract `Actions de Recherche Concert\'{e}es' no. 04/09-312) and by the Federal Ministry of Education and Research (BMBF) through the program `Spitzenforschung und Innovation in den Neuen L\"{a}nden' (contract `Potsdam Research Cluster for Georisk Analysis, Environmental Change and Sustainability' D.1.1).
Ch. A. was supported by the PAI 2007-2011 `NOSY-Nonlinear systems,
stochastic processes and statistical mechanics' (FD9024CU1341) contract
of ULB. The work of V. B. is partially supported by the European Space
Agency contract No. ESA AO-2004-070.

\bibliographystyle{elsarticle-num}
\bibliography{GCAB2011.bib}

\end{document}